\documentclass[12pt, a4paper]{article}

\usepackage{amsmath}
\usepackage{amssymb}
\usepackage{amsfonts}
\usepackage{multirow}
\usepackage{amsthm}
\usepackage{color}
\usepackage{natbib}
\usepackage{graphicx}
\usepackage{moreverb}
\usepackage[colorlinks,bookmarksopen,bookmarksnumbered,citecolor=red,urlcolor=red]{hyperref}
\usepackage{natbib}

\begin{document}

\title{Bayesian models for data missing not at random in health examination surveys}

\author{Juho Kopra\\
Department of Mathematics and Statistics,\\ University of Jyv\"askyl\"a, Finland\\[2mm]
and \\[2mm]
Juha Karvanen \\
Department of Mathematics and Statistics,\\ University of Jyv\"askyl\"a, Finland\\[2mm]
and \\[2mm]
Tommi H\"ark\"anen\\
National Institute for Health and Welfare,\\ Helsinki, Finland	
}
\date{20th June 2017 (Draft version)}

\maketitle

\begin{abstract}
In epidemiological surveys, data missing not at random (MNAR) due to survey nonresponse may potentially lead to a bias in the risk factor estimates. We propose an approach based on Bayesian data augmentation and survival modelling to reduce the nonresponse bias. The approach requires additional information based on follow-up data. We present a case study of smoking prevalence using FINRISK data collected between 1972 and 2007 with a follow-up to the end of 2012 and compare it to other commonly applied missing at random (MAR) imputation approaches. A simulation experiment is carried out to study the validity of the approaches. Our approach appears to reduce the nonresponse bias substantially, where as MAR imputation was not successful in bias reduction.
\end{abstract}

\newpage
\section{Introduction}
Population level estimates of risk factors are of major interest in epidemiology. Data on risk factors such as blood pressure, cholesterol level, body mass index, alcohol consumption and daily smoking are often collected in health examination surveys (HES). In a HES, the data on risk factors are gathered usually via both questionnaires and physical measurements. The trends of population level risk factors are monitored and they are valuable input for policy decisions.

Missing data by unit nonresponse occurs in a HES as invitees neither participate to physical measurements nor return a survey questionnaire. The decision about participation have been found to depend on the risk factors, such as smoking \citep{shahar1996}, either directly or via a common cause such as health awareness. This may be deduced from the fact that the non-participants have a higher risk of death \citep{jousilahti2005,harald2007,karvanen2016}. This dependence causes missing data to be classified as missing not at random (MNAR) \citep{rubin1976}. Because the data are MNAR, the population level risk factors calculated from the participants' data are biased, and they usually give an overly healthy view of the population. Biased estimates of risk factor prevalence may seriously misinform decision makers. Instead of analysing only the participants data, the posterior distributions of risk factor levels of a whole sample, including non-participants, should be estimated. This requires external information and modelling assumptions.

In this paper, we demonstrate how follow-up data on endpoints associated with the risk factor of the interest provides external information that allows us to reduce the bias caused by selective non-participation. We propose a Bayesian method for the estimation of risk factor prevalence and the missing data mechanism, when the data are MNAR. 

Our datasets origin from the FINRISK studies, which are national HES providing information about the health of Finns. 
We improve and extend an earlier work \citep{kopra2015} on the estimation of smoking prevalence from FINRISK data. The key improvements are:
a fully Bayesian model is used,
the survival model is more flexible, and
informative prior is utilised instead of assumption of conditional independence \citep[Eq. (2)]{kopra2015}. 
Differently from \cite{kopra2015} the study years 2002 and 2007 are included in the modelling.

Next section describes the data of the FINRISK studies and follow-up. Section \ref{section-model} presents the Bayesian model and the priors that we apply to smoking prevalence estimation. Section \ref{section-model-fitting} explains model fitting, and Section \ref{section-simulation} provides a simulation study on the proposed approach. We evaluate alternative methods in Section \ref{section-alternative-methods}. In Section \ref{section-truedata} we apply our approach to real data from the FINRISK studies and provide smoking prevalence estimates for both men and women.
Section \ref{section-discussion} discusses the results and methods presented.

\section{Data description}
Our HES data contain eight FINRISK studies conducted in selected geographical areas of Finland once in every five years in 1972-2007 \citep{laatikainen2003,harald2007}. 
In each study year, persons were selected to the FINRISK studies in a random sampling stratified by region, gender and 10-year age group. 
Our data are restricted to the two regions (Northern Savonia and North Karelia) that have been included in all eight studies. 
In total, the data contain $52,325$ persons including $9,928$ persons with missing smoking indicator.

Each person selected to the study received a letter of invitation, in which he or she was asked to fill in a survey questionnaire and participate to physical measurements in the local survey site. If the person participated, the filled questionnaire was collected and the physical measures were taken. If the person did not participate, then risk factors are missing, but background variables, study year, age, gender, and region, are known from the sampling frame. Table \ref{table-participation} shows that the participation rates have dramatically decreased from 1972 to 2007. It can be also seen that women have participated more actively than men in all study years. We also know that person's age affects participation \citep{kopra2015}.

\begin{table}[!htbp] \centering 
	\caption{Participation rates (\%) and size of survey sample ($n$), by gender, region and year. The participation rates of 1972 and 1977 are approximated (*) as the region information of non-participants is missing for these years.} 
	\label{table-participation} 
	\begin{tabular}{@{\extracolsep{5pt}} cccccc} 
		\\[-1.8ex]\hline 
		\hline \\[-1.8ex] 
		&	& \multicolumn{2}{c}{ North Karelia } & \multicolumn{2}{c}{ Northern Savonia } \\
		Year	& & men & women & men & women \\ 
		\hline \\[-1.5ex] 
		1972	&\% & 84.3* & 88.5* & 88.4* & 91.3* \\ 
		&$n$	& 2,641 & 2,607 & 3,574 & 3,555 \\[2mm] 
		1977	&\% & 85.7* & 89.0* & 90.1* & 93.0* \\ 
		&$n$	& 2,323 & 2,382 & 3,223 & 3,391 \\[2mm] 
		1982	&\% & 76.1 & 83.2 & 80.8 & 86.0 \\ 
		&$n$	& 2,007 & 2,019 & 1,810 & 1,566 \\[2mm] 
		1987	&\%	& 78.8 & 85.3 & 80.3 & 86.3 \\ 
		&$n$	& 1,971 & 1,976 & 979 & 988 \\[2mm] 
		1992	&\%	& 68.2 & 80.8 & 75.9 & 83.8 \\ 
		&$n$	& 984 & 993 & 982 & 990 \\[2mm] 
		1997	&\%	& 72.1 & 75.3 & 70.8 & 79.8 \\ 
		&$n$	& 1,052 & 1,020 & 990 & 997 \\[2mm] 
		2002	&\%	& 66.5 & 76.2 & 66.2 & 78.2 \\ 
		&$n$	& 1,021 & 1,011 & 1,000 & 1,000 \\[2mm] 
		2007	&\%	& 63.0 & 71.9 & 61.1 & 70.4 \\ 
		&$n$	& 811 & 825 & 817 & 820 \\
		\hline \\[-1.8ex] 
		Total &\% & 77.5 & 83.5 & 81.5 & 87.1 \\
		&$n$ & 12,810 & 12,833 & 13,375 & 13,307
	\end{tabular} 
\end{table} 

Our HES data were linked together with follow-up data of all participants and non-participants. The follow-up data contains the exact dates and diagnoses (ICD codes) of hospitalisations and deaths. In Finland, this kind of follow-up data can be collected from administrative registers for both participants and non-participants. The follow-up period started at the time of study for each person and ended on 31st December 2012 for all FINRISK study years. Thus, the length of the follow-up period varies by study years.

It is well-known that smoking is a key risk factor for lung cancer and chronic obstructive pulmonary disease (COPD) \citep{doll1956,mannino2007}. Thus, we use lung cancer and COPD events together as an endpoint. Table \ref{table1} 
shows that non-participants have a higher rate of disease events than participants.

\begin{table}[!htbp] \centering 
	\caption{The total count of observed lung cancer and COPD events, events per $1000$ follow-up person-years, and participation rate by region and gender.}
	\label{table1}
	\resizebox{\textwidth}{!}{\begin{tabular}{@{\extracolsep{1pt}} cllrcc}
		\\[-1.8ex]\hline 
		\hline \\[-1.5ex]
		Region & Gender & Participant & Events & Events/1000 years & Participation (\%) \\ 
		\hline \\[-3mm]
		N. Karelia & Men   & Yes & $387$ & $1.75$ & \multirow{2}{*}{\vspace{1mm}$77.4$}\\ 
		N. Karelia & Men   & No  & $166$ & $3.14$ & \\ 
		\hline\\[-3mm]
		N. Savonia & Men   & Yes & $479$ & $1.85$ & \multirow{2}{*}{\vspace{1mm}$81.6$}\\ 
		N. Savonia & Men   & No  & $129$ & $2.85$ & \\ 
		\hline\\[-3mm]
		N. Karelia & Women & Yes & $75$  & $0.28$ & \multirow{2}{*}{\vspace{1mm}$83.6$}\\ 
		N. Karelia & Women & No  & $43$  & $1.02$ & \\ 
		\hline\\[-3mm]
		N. Savonia & Women & Yes & $62$  & $0.21$ & \multirow{2}{*}{\vspace{1mm}$87.0$}\\
		N. Savonia & Women & No  & $33$  & $0.94$ & \\ 
		\hline \\ 
	\end{tabular}}
\end{table} 

We limit in our analysis the age range to 25--64 years-old and select the subset of healthy persons with respect to our endpoint variables. The two exceptions are 1972  and 1977 studies, which have age ranges of 25--59 and 30--64 years-old, respectively. 

\section{Bayesian model}
\label{section-model}
The modelling is based on the idea that although it is impossible to directly observe the smoking status of non-participants, we can obtain information on smoking indirectly from the follow-up data. More precisely, the modelling uses 
the observed incidence differences of the smoking-based diseases between participants and non-participants, which allows us to adjust the estimates of smoking prevalence.
 Full Bayesian approach is applied, and model fitting is executed using Markov chain Monte carlo (MCMC) methods \citep{robertandcasella2004}.

\subsection{Notation for the data}
We introduce our model using causal models with design \citep{karvanen2015}, and make a difference between measurements and underlying causal variables. The model is presented in Figure  \ref{figure-dag}. For each person $i=1,\dots,N$ invited to the survey, we denote participation indicator by $M_i$, which takes the value $M_i=1$ if person $i$ participated, and value $M_i=0$ otherwise. Value $M_i=0$ indicates missing risk factor data. The indicator of self-reported daily smoking is denoted by $Y_i$ and the corresponding measurement by $Y_i^*$. Variable $Y_i$ takes value 1, if a person is a daily smoker, and $0$ otherwise. Class $Y_i=0$ includes earlier smokers who quitted. The value of $Y^*_i$ is known for the participants, then $Y^*_i=Y_i$, but missing for the non-participants. We denote by vector $X_i^*$ the variables age $a_i$, gender $g_i$, region $r_i$ and study year $s_i$ in the background data observed for all sample members. The values $g_i=0$ stand for men, and $g_i=1$ for women. The North Karelia region is denoted by $r_i=0$ and Northern Savonia by $r_i=1$.

We denote $T_i$ as the age at the day of diagnosis, which may also be the age at the time of death if a person dies without previous lung cancer or COPD diagnoses and the death is caused by either of the two diseases. If the person has not been diagnosed, the corresponding measurement $T^*_i$ is missing, and $T_i$ is right-censored. Variable $T_{\text{cens},i}$ is the age of the person $i$ in the end of the year 2012, which is the end of our follow-up period, or the age of death for the person who have died before the end of the year 2012. The variable $T_{\text{obs},i}$ is the minimum of $T_i$ and $T_{\text{cens},i}$, so $T_{\text{obs},i}=\min(T_i,T_{\text{cens},i})$, and $T_{\text{cens},i}^* = T_{\text{cens},i}$ and $T_{\text{obs},i}^*=T_{\text{obs},i}$.

\subsection{Submodels}
\begin{figure}[!htbp]
	\center
	\includegraphics[width=1.0\textwidth]{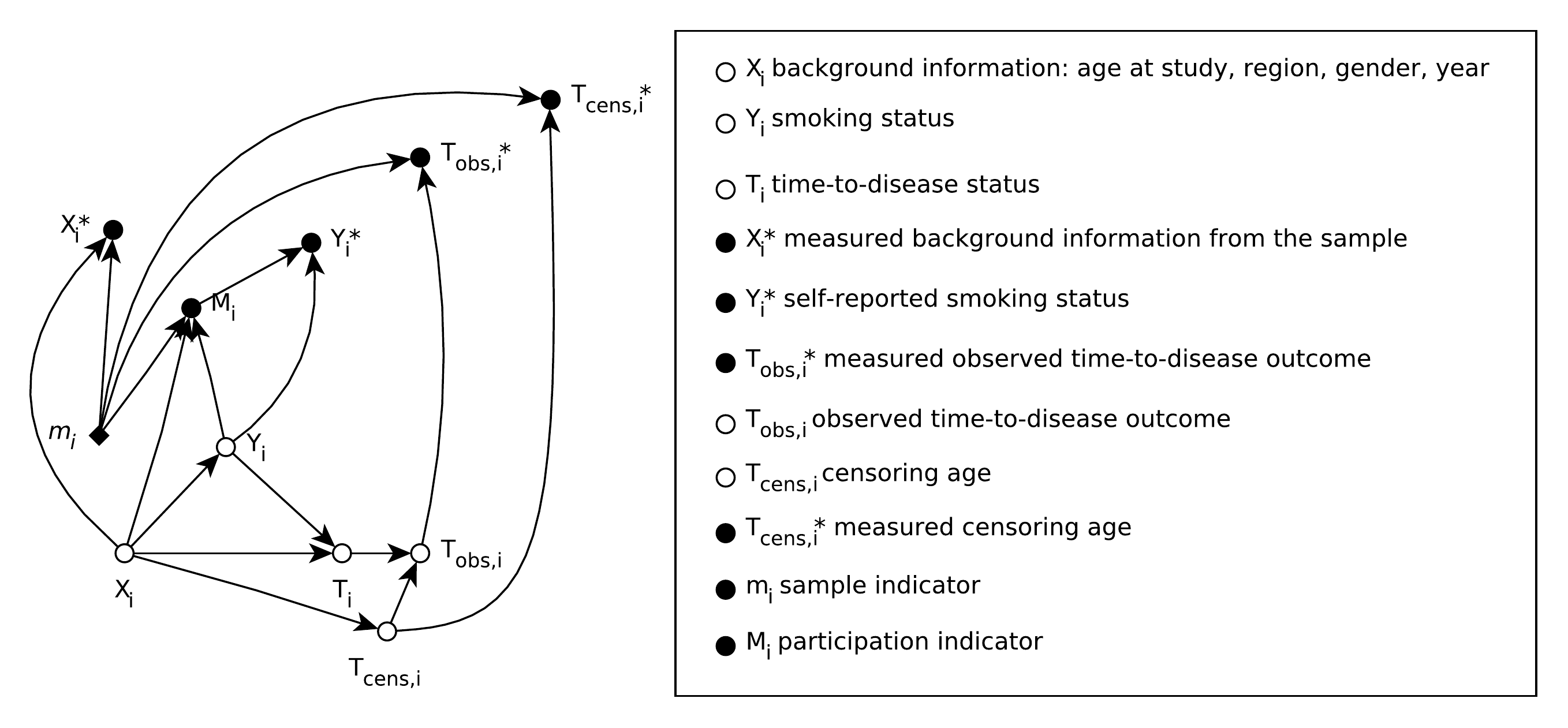}
	\caption{A graph representing the model and the dependencies between the variables of HES data and the follow-up data. Direct causal effects are represented as arrows. 
	Measurement variables are denoted with asterisk, e.g. $X_i^*$, and are presented as filled circles. The causal variables do not have asterisk symbol (e.g. $X_i$), and they are drawn unfilled to indicate that they are not observed directly but via measurement variables. 
	The measurement variables always have one participation indicator ($m_i$ or $M_i$) and one causal variable as their parent. The graph tells that $X_i^*$, $T_{\text{obs},i}^*$ and $T_{\text{obs},i}^*$ are collected for each member of the sample while $Y_{i}^*$ is measured only for participants, and is missing for the non-participants.
	}
	\label{figure-dag}
\end{figure}

The joint model for data from a HES linked with follow-up consists of three submodels:
\begin{enumerate}
	\item a participation model in which participation is explained by daily smoking and background variables (arrow $X_i \rightarrow M_i$ in Figure \ref{figure-dag}),
	
	\item a risk factor model for daily smoking given the background variables (arrow $X_i \rightarrow Y_i$),
	
	\item a survival model for the follow-up data given the daily smoking and background variables (arrows $Y_i \rightarrow T_i$ and $X_i \rightarrow T_i$).
\end{enumerate}
These three submodels together form a joint model for the data, which we call Bayesian MNAR model, see Figure \ref{figure-dag}. 
The arrows $X_i \rightarrow M_i$ and $Y_i \rightarrow M_i$ correspond to the participation submodel, that can be written as $P(M_i=1|X_i,M_i)$. The arrow $X_i \rightarrow Y_i$ corresponds to the risk factor submodel (distribution $P(Y_i|X_i)$), and the arrows $Y_i \rightarrow T_i$ and $X_i \rightarrow T_i$ correspond to the survival model (distribution $P(T_i|Y_i,X_i)$). All the submodels are fitted together because each of them contains the indicator of smoking, which has missing values to be imputed.

\subsection{Participation model}
\label{subsection-participation-model}
First, our model for participation indicator $M_i$ is
\begin{equation}
\text{logit}(P(M_i=1|X_i,Y_i)) = \alpha_{0[g_i,s_i]} + \eta_{[g_i,s_i]}Y_i + \alpha_{1[g_i,Y_i]}(a_i-45) + \alpha_{2}r_i,\label{eq3.1}
\end{equation}
$g_i,s_i,a_i$ and $r_i$ are part of $X_i$ and they stand for gender, study year, age, and region, respectively. Variable $Y_i$ stands for smoking. The roles of model parameters $\alpha_{0[g_i,s_i]}$, $\eta_{[g_i,s_i]}$, $\alpha_{1[g_i,Y_i]}$ and $\alpha_{2}$ are explained below. Parameter $\alpha_{0[g_i,s_i]}$ is a regression coefficient (intercept) which varies over the levels of $g_i$ and $s_i, i = 1,\dots,N$. The variable $g_i$ is binary and $s_i$ has eight possible values, which create the total of 16 intercept parameters. The parameters $\eta_{[g_i,s_i]}$ are gender-specific regression coefficients modelling how daily smoking affects participation in each year. We also take into account how the age of person affects participation; the gender-specific coefficients $\alpha_{1[g_i,0]}$ and $\alpha_{1[g_i,1]}$ model how age affects participation for non-smokers and smokers, respectively. 
The parameter $\alpha_{2}$ describes the differences in participation between the two regions.

\subsection{Risk factor model}
\label{subsection-risk-factor-model}
Next, we need to model smoking indicator $Y_i$ by background variables $X_i = (g_i,s_i,a_i,r_i)$. We use a logistic regression model
\begin{equation}
\text{logit}(P(Y_i=1|X_i)) = \beta_{0[g_i,r_i,s_i]} + (s_i - a_i - 1938)\beta_{1[g_i,r_i,s_i]},
\label{eq3.2}
\end{equation}
where coefficients $\beta_{0}$ and $\beta_{1}$ vary between groups defined by combinations of gender $g_i$, region $r_i$ and study year $s_i$ similarly as in \eqref{eq3.1}. The year of birth $s_i - a_i$ for person $i$ is centered at its rounded population average 1938 in the model.

\subsection{Survival model}
\label{subsection-survival-model}
To define a survival model for $P(T_i|X_i,Y_i)$, a counting process notation is used. Let $N_i(t)$ stand for the count of disease diagnoses up to age $t$ for person $i$. Let $dN_i(t)$ be the increment of the counting process over one-year time interval $[t,t+1)$, and let $t$ take discrete values $25,26,\dots,100$. Now, we model $dN_i(t)$ with a piecewise constant hazard model assuming that for each one-year time-period, the gender-specific hazard $h_{0,g}(t)$ remains constant ($g=0,1$ stands for the gender). The model for follow-up data is
\begin{align}
&dN_i(t) \sim \text{Poisson}(\lambda_i(t))\\
&\lambda_i(t) = \begin{cases}
\exp\left(\gamma_{1}Y_i\right) h_{0,0}(t), &\hbox{ given that } T_i\ge t \hbox{ and }g=0\\
\exp\left(\gamma_{2}Y_i\right) h_{0,1}(t), &\hbox{ given that } T_i\ge t \hbox{ and } g=1\\
0, & T_i< t,
\end{cases}
\label{eq3.3}
\end{align}
where $\gamma_1$ and $\gamma_2$ model how smoking increases the hazard for men and women, respectively.

\subsection{Prior distributions}
For the participation model, we use informative prior distribution for the difference between smokers and non-smokers. An informative prior for $\eta_{[g_i,s_i]}$ is derived as follows. We consider a 45-year-old non-smoker, who participates with probability $p=0.7$, and elicit the corresponding prior probability for a smoker, who is otherwise similar. We consider that there is a 15\% chance that the participation prior probability $p$ is less than 0.50, about 30\% chance for less than 0.60, and 50\% chance for less than 0.70. These considerations together with an assumption on a logistic distribution for $\eta_{[g_i,s_i]}$ lead to prior distribution
\begin{equation*}
\eta_{[g_i,s_i]} \sim \text{Logistic}(\mu=0,s=2.05^{-1}), 
\end{equation*}
which makes the prior distribution of $p$ to have expected value $E(p)=0.676$ and 95\% credible interval $[0.281, 0.933]$. Here, logistic distribution density function is $$f_{\text{logistic}}(x|\mu,s) = \frac{e^{(x-\mu)/s}}{s(1+e^{(x-\mu)/s})^2},$$ for $x,\mu \in \mathbb{R}$ and scale parameter $s > 0$. 
%

The prior distributions for participation model coefficients $\alpha_{0[g_i,s_i]}$ and $\alpha_{1[g_i,Y_i]}$, and risk factor model parameters $\beta_{0[g_i,r_i,s_i]}$ and $\beta_{1[g_i,r_i,s_i]}$ are normal distributions with mean $\mu=0$ and variance $\sigma^2=1000$ (uninformative priors).

Survival model parameters $\gamma_{1}$ and $\gamma_{2}$ are also a priori normally distributed with $\mu=0$ and $\sigma^2=1000$. Our prior distribution for baseline hazard $h_{0,g}(t)$ is monotonically increasing with age
\begin{align*}
&h_{0,g}(25) \sim \text{Uniform}(0,20)\\
&h_{0,g}(t) \sim \text{Uniform}(h_{0,g}(t-1),20),\hspace{1.4cm} \text{where } t=26,27,\dots,100,
\end{align*}
where $g$ stands for gender, $0$ for men and $1$ for women. This means that model assumes that risk of smoking-based diseases only increase with age. This assumption seems to be in agreement with our data.

\section{Model fitting}
\label{section-model-fitting}
As the number of model parameters (316) and missing values (9,928) is large, there are over 10,000 variables to sample at each iteration of the MCMC model fitting process.
This creates a computational challenge for Bayesian model fitting. 
The Markov chains typically require thousands of iterations or more to obtain satisfactory convergence, which requires a lot of computing time.

To impute the missing values for smoking indicators, the data augmentation was applied \citep{tannerwong1987}. A Bayesian MNAR model described in Figure \ref{figure-dag} and Sections \ref{subsection-participation-model}, \ref{subsection-risk-factor-model} and \ref{subsection-survival-model} was used. The augmented data for smoking indicator $Y_i$ are drawn from fully conditional distribution $P(Y_i|M_i=0,X_i,T_i)$, given its parent nodes ($X_i$) and child nodes ($M_i$ and $T_i$).

We used Just Another Gibbs Sampler -software (JAGS) \citep{plummer2003}, R \citep{r2016} and {\tt rjags} package \citep{rjags} to fit the model. Seven parallel MCMC chains were used. Each chain had $9000$ burn-in iterations, $45900$ actual iterations with thinning interval $75$, which makes a total of $612$ iterations per chain to be recorded. The time consumed for this model fitting using parallel chains was about $107$ hours, which makes $7$ seconds per each iteration and less than $0.7$ milliseconds per each parameter for one iteration. 
The high absolute number of missing values ($9,928$) explains the long running time.
Because of high autocorrelations in the chains, we decided to use a long thinning interval and many iterations to reduce the autocorrelations in the saved iterations of the MCMC and larger sample of the posterior for the final trends. The convergence of the chains was examined both visually and using Brooks-Gelman $\hat{R}$-diagnostics \citep{brooks1998rhat}. All the $\hat{R}$ test statistics for model coefficients were below $1.01$ which indicates convergence.

\section{Simulation study}
\label{section-simulation}
We carried out a simulation experiment to demonstrate the performance of the model. The simulated data allow us to compare the performance of the estimated prevalence of smoking with true prevalence, which is known with the simulated data but not with the HES data. The actual values of the background variables from the FINRISK study were used together with parameter estimates from a preliminary analysis. Thus, the simulation experiment had conditions similar to the real data, e.g. the smoking prevalence had a decreasing trend for men and an increasing trend for women. For both genders the participation was selective, and the participation rate had a decreasing trend.

The simulation was implemented using R language, and the data were simulated from the Bayesian MNAR model presented in Section \ref{section-model}. Because of the computational burden of model fitting, the model was fitted into a single simulated data set.

The model fitting for the Bayesian MNAR model with simulated data was implemented as described in Section \ref{section-model-fitting}. All the $\hat{R}$-diagnostics were below $1.01$ which indicates convergence. We inspected the posterior correlations between the model parameters and found strong correlations ($\ge 0.9$ or $\le -0.9$) between some of the parameters. 
In the risk factor model, the strongest correlations were observed between the parameters $\beta_{0}$ and $\beta_{1}$. The median of these correlations was $-0.434$ and the range was $[-0.905,0.661]$. Conditioning with $M_i$ likely causes these posterior correlations. On the participation model, the strongest posterior correlations were found between the $\alpha_{0[g_i,s_i]}$ and $\eta_{[g_i,s_i]}$. For men, those eight correlations ranged between $-0.972$ and $-0.899$, and for women between $-0.860$  and $-0.665$. In the survival model, strong positive correlations occurred particularly between the hazards of consecutive years, which is natural to this type of models. The highest correlations for men were $0.920$ and for women $0.952$.

\begin{figure}
	\centering
	\includegraphics[width=1.0\textwidth]{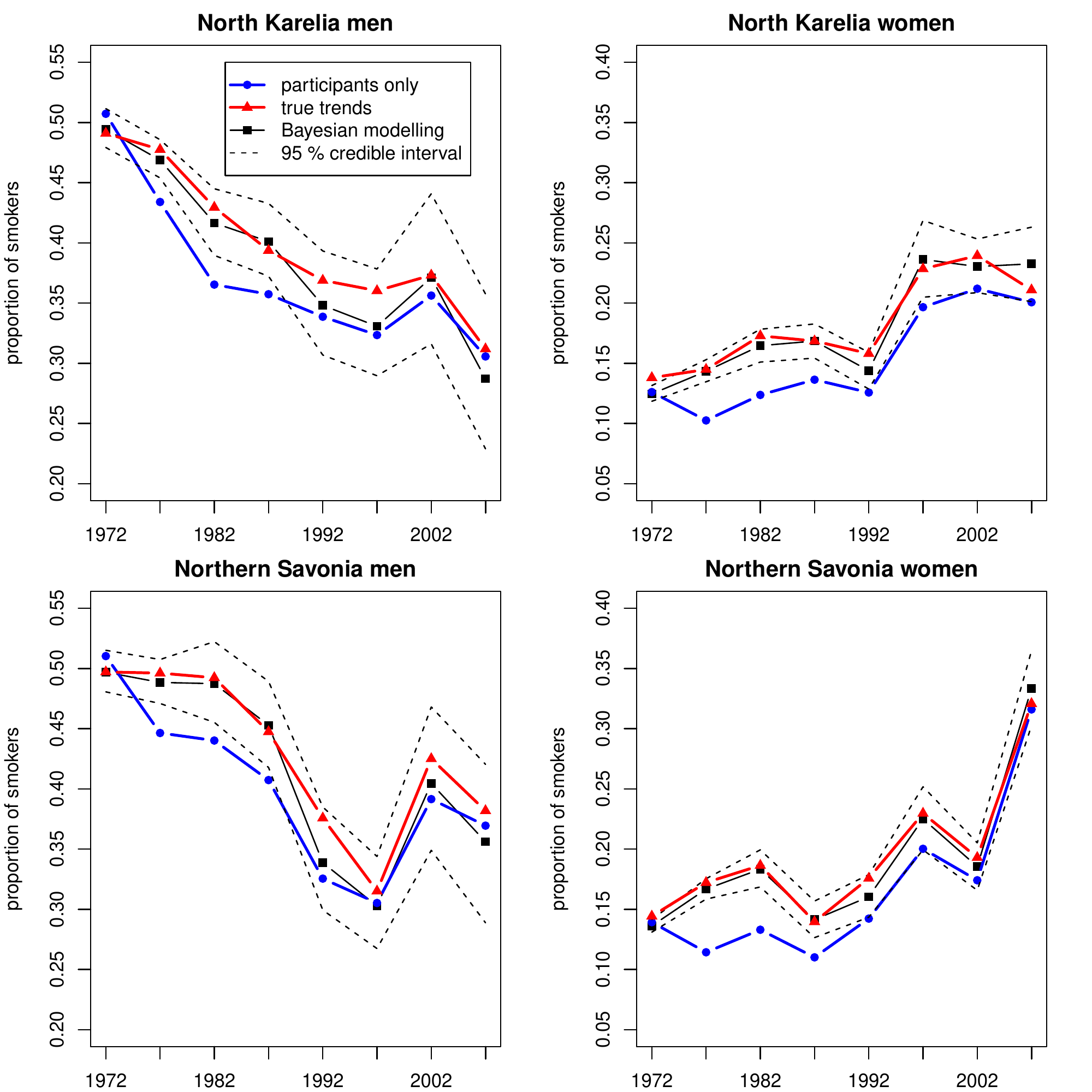}
	\caption{Trends for the simulation experiment. The red line with triangles is the true proportion of smokers used in the simulation. The blue line with circles is estimated from participants only and the black line with squares is model-based posterior mean with 95\% credible intervals (dashed black line) calculated from the simulated data.}
	\label{simulation_trends}
\end{figure}

It can be seen from Figure \ref{simulation_trends} that with an exception of women in 1972, the true prevalence is located inside the credible interval and there is no indication of systematic bias in the posterior mean. In contrast, the prevalence calculated using only the participants systematically underestimates the true prevalence. As the same family of models was used both to simulate the data and to fit the parameters, the Bayesian MNAR approach is expected to perform very well. However, the experiment demonstrates that the model can be estimated from the data.

\section{Alternative methods}
\label{section-alternative-methods}
In addition to the Bayesian MNAR approach, we considered two alternative modelling approaches and the complete case analysis. Both modelling approaches utilize the MAR assumption and the complete case analysis uses data on participants only. In terms of Figure \ref{figure-dag}, the MAR assumption omits the arrow $Y_i \rightarrow M_i$, which means that participation is not selective with respect to smoking. 

First, the Bayesian MAR approach differs from the Bayesian MNAR approach such that the entire survival model \eqref{eq3.3} is omitted, and the regression coefficient $\eta_{[g_i,s_i]}$ is fixed to zero in participation model \eqref{eq3.1}. 

We also used the frequentist MAR model which was implemented using the \texttt{mice} package in R \citep{mice}. The missing smoking indicator was imputed using a logistic regression model that had full interactions between year of birth, gender, region and year, and full interactions between gender, event indicator, and age at the event/censoring. The year of birth and the age at event/censoring were used as linear covariates and the other variables were categorical.

We fitted these alternative approaches to data simulated from the MNAR model which is selective with respect to smoking. The trend estimates are presented in Appendix A in Table \ref{table4}. We calculated root mean square errors (RMSE) for the Bayesian MNAR model and each of the alternative approaches. The RMSE was calculated using formula 
\begin{equation}
\text{RMSE}(y,y_{\text{true}}) = \sqrt{\frac{1}{n}\sum_{i=1}^n (y_i-y_{\text{true},i})^2},\nonumber
\end{equation}
where $y$ is the estimated smoking prevalence (\%) and $y_{\text{true}}$ is the true smoking prevalence from simulation. 
The RMSE was calculated over the regions, the genders and the study years, which gives one RMSE value for each approach.

The Bayesian MNAR approach had the smallest RMSE, $1.65$. The Bayesian MAR and the frequentist MAR methods have very similar RMSE with each other, $3.34$ and $3.37$, respectively. These two methods were slightly more accurate than the complete case approach with RMSE $3.48$.

\section{Application to FINRISK data}
\label{section-truedata}
\begin{figure}
	\centering
	\includegraphics[width=1.0\textwidth]{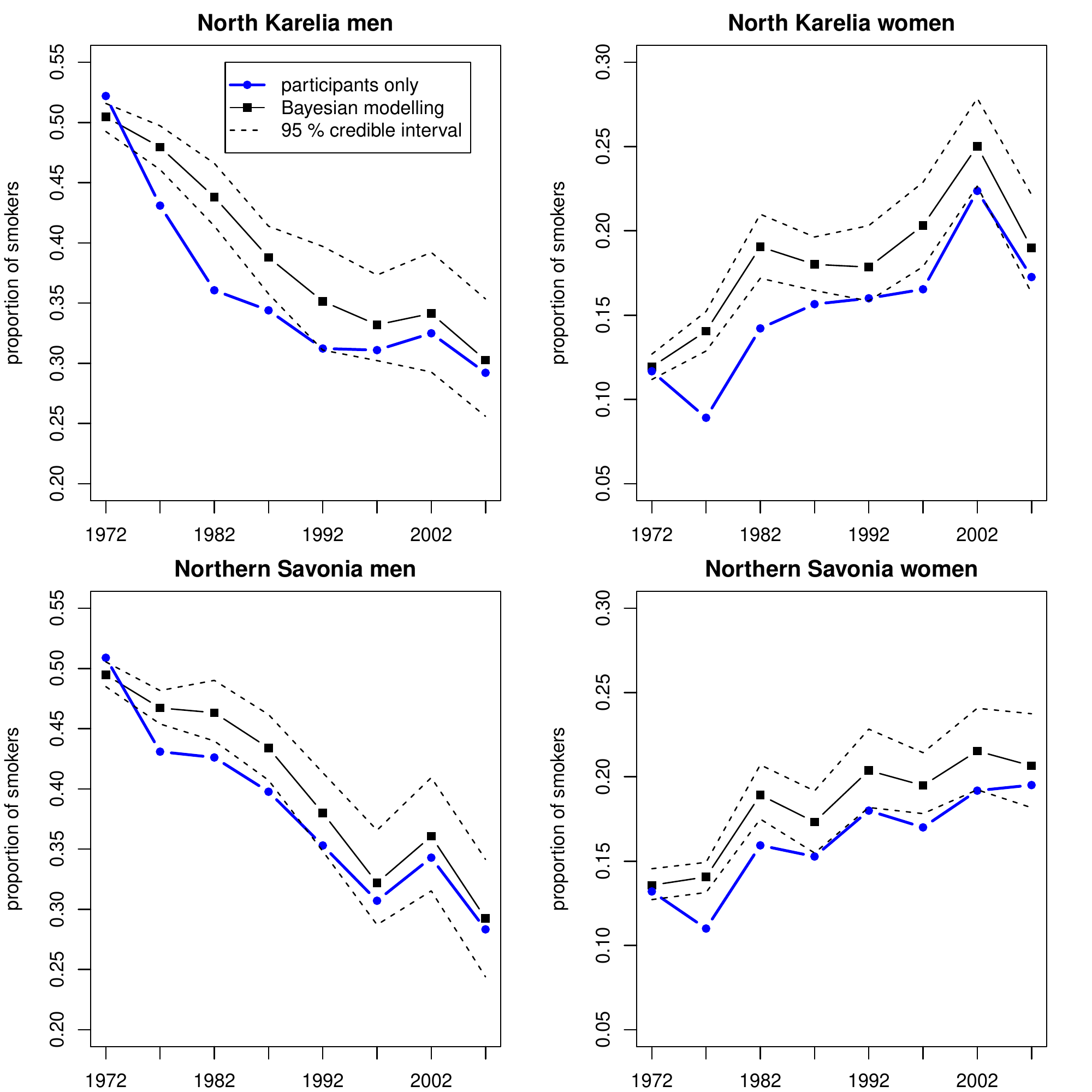}
	\caption{Participant trends (blue line with circles) and model-based posterior trends (black line with squares) with 95\% credible intervals (dashed black line) for the FINRISK data. For numeric presentation of trends, see Table \ref{table4} in Appendix A.}
	\label{finrisk_trends}
\end{figure}

The trends of daily smoking for the FINRISK data estimated using the Bayesian MNAR model are reported in Figure \ref{finrisk_trends} and compared to participant trends, which are often reported in HES. 
The difference between the smoking prevalence estimates of the complete case (participants) and the Bayesian MNAR approaches is the highest for the study years 1977, 1982, and 1987 (Figure \ref{finrisk_trends} and Table \ref{table3} in Appendix A). Starting from 1992, the complete case trends are within the 95\% credible interval of the Bayesian MNAR model, but they are systematically below the posterior mean. 
The proportion of missing data is higher for the later study years than the earlier ones, which makes the credible intervals wider.

The values of the region variable $r_i$ for the study years 1972 and 1977 were missing for non-participants. We used a single imputation with fixed probabilities\\ \mbox{$P(r_i=1|s_i=1972)=0.495$} and $P(r_i=1|s_i=1977)=0.493$ as in \cite{kopra2015}. We decided to use single imputation a prior to model fitting because the regions seemed to be rather similar with respect to smoking prevalence, and the participation rates were high.

We executed sensitivity analysis to find out if model can be fitted with even less informative prior. We tried prior distributions for $\eta$ with $s$ parameter set to $(2.05/2)^{-1}$, which corresponds to doubling the prior variance. We found out that the Markov chains do not converge. More precisely, convergence problems were found with $\eta$-parameters for which the $\hat{R}$s ranged between $1.097$ and $1.828$ after $45900$ iterations and $9000$ burn-in. Thus, it appears that vague prior distributions are not applicable.


\section{Discussion}
\label{section-discussion}
We have proposed the Bayesian MNAR modelling approach to reduce the non-participation bias and applied the approach to the FINRISK studies. 
In a simulation experiment, we compared our approach to the Bayesian MAR approach, to the complete case analysis, and the frequentist MAR imputation. The latter two were easier to use than the Bayesian MNAR approach but did not substantially reduce the non-participation bias. The proposed approach appears to reduce the non-participation bias.

Trends by the Bayesian and the frequentist MAR approaches are essentially the same as participants' trends. Thus, the MAR approaches do not reduce the bias of risk factor levels by much. Although there may be ways to improve the imputation model \citep{whiteroyston2009}, the MAR imputation do not account for selective non-participation.

The information about non-participants' risk factor levels comes from the survival data. The risk factor of interest must be a strong predictor of the survival outcome. The estimation of survival model parameters requires that a sufficient number of events have been recorded. This implies that the length of follow-up must be long enough and if the event is rarely observed, the number of persons must be high.  The information obtained from the survival model may be insufficient by itself and need to be supported by an informative prior on the selection mechanism.

In many countries, it is not technically or legally possible to link the HES data of non-participants to follow-up data. The requirement on the availability of survival data for both participants and non-participants is a major limitation for the proposed approach.

The Bayesian model fitting is often a computational challenge, particularly when the amount of missing data is large. The memory management and computation time were issues in our case study, due to a large amount of missing data. In our first attempts, we tried to save all the imputations, which filled the RAM memory of the computer (16 GB) quite rapidly. We later realised that it is possible to calculate sufficient (summary) statistics, e.g. count of smokers and non-smokers by gender, year and region, and store only them. We also reduced the time required per one iteration by coarsening the continuous covariates (the age) of the survival model and summing over discrete or discretised covariates, and by modelling the number of events in each risk group using Poisson distribution.

Another challenge was the posterior correlations caused by the model structure and non-random missingness. One possibility to alleviate this problem could be to develop a custom MCMC algorithm with blocked updating of the parameters with high posterior correlations \citep{haario2001}. However, this often requires custom programming and is therefore much more laborious than an application based on JAGS, which we have applied.
Thus, if one needs to use data with more missing data than in our case study  or multiple variables with MNAR missingness, we recommend using some specialized MCMC algorithms, or possibly the iterated importance sampling algorithm \citep{celeux2006}.

Additional fully observed covariates can be added into the model without excessive increase in computational burden. If more variables with missing data are imputed, the model fitting is slowed down proportional to increase of absolute amount of missing values. This is because at each MCMC iteration all the missing values need to be imputed.

Selective non-participation in HES is an important problem that may have implications to the decisions on health policy. Our solution is not simple to implement but the reduction of selection bias makes it worth of the effort.

\section*{Acknowledgements}
The work was supported by the Academy of Finland [grant number 266251].

\bibliographystyle{wb_stat}
\bibliography{references}

\newpage
\section*{Appendix A}
\begin{table}[!htbp] \centering
	\footnotesize 
	\caption{The trends and 95 \% credible intervals of simulation experiment and alternative approaches. } 
	\label{table3} 
	\renewcommand{\arraystretch}{0.6}
	\resizebox{\textwidth}{!}{\begin{tabular}{@{\extracolsep{-2pt}} clllll} 
		&  & \multicolumn{2}{c}{Northern Karelia} & \multicolumn{2}{c}{North Savonia} \\\hline
		year & method & men & women & men & women \\ 
		\hline \\[-1.5ex] 
$1972$ & Bayes+MAR & 50.9 (50.2, 51.7) & 50.3 (49.6, 51.0) & 12.2 (11.8, 12.7) & 14.2 (13.7, 14.6) \\ 
$1972$ & Bayes+MNAR & 49.5 (47.8, 51.2) & 48.8 (47.3, 50.5) & 12.4 (11.7, 13.2) & 14.7 (14.0, 15.5) \\ 
$1972$ & Complete case & 50.6 (48.5, 52.7) & 50.0 (48.2, 51.7) & 12.2 (10.8, 13.5) & 14.2 (13.0, 15.4) \\ 
$1972$ & mice & 51.0 (49.0, 53.0) & 50.4 (48.6, 52.1) & 12.2 (10.8, 13.6) & 14.2 (13.0, 15.5) \\ 
$1972$ & True & 49.1              & 48.9              & 12.9              & 15.2              \\ \hline
$1977$ & Bayes+MAR & 42.6 (41.8, 43.4) & 42.9 (42.2, 43.7) & 8.8 (8.5, 9.2) & 11.9 (11.5, 12.3) \\ 
$1977$ & Bayes+MNAR & 46.9 (45.7, 48.3) & 47.3 (45.8, 48.8) & 12.2 (11.2, 13.2) & 16.6 (15.7, 17.7) \\ 
$1977$ & Complete case & 42.6 (40.5, 44.8) & 43.0 (41.1, 44.8) & 8.9 (7.7, 10.1) & 11.9 (10.8, 13.1) \\ 
$1977$ & mice & 42.8 (40.7, 45.0) & 43.1 (41.3, 44.9) & 8.8 (7.4, 10.3) & 11.9 (10.6, 13.2) \\ 
$1977$ & True & 47.0              & 47.5              & 13.2              & 17.5              \\ \hline
$1982$ & Bayes+MAR & 38.4 (37.2, 39.5) & 42.0 (40.6, 43.3) & 14.5 (13.9, 15.2) & 13.1 (12.3, 13.9) \\ 
$1982$ & Bayes+MNAR & 44.0 (41.4, 46.5) & 46.7 (43.5, 50.2) & 18.3 (16.8, 19.9) & 17.7 (16.0, 19.4) \\ 
$1982$ & Complete case & 38.2 (35.8, 40.6) & 41.7 (39.0, 44.4) & 14.5 (12.8, 16.2) & 13.0 (11.2, 15.0) \\ 
$1982$ & mice & 38.6 (36.2, 41.1) & 42.0 (39.5, 44.6) & 14.4 (12.7, 16.0) & 13.0 (11.1, 15.0) \\ 
$1982$ & True & 45.0              & 47.7              & 19.6              & 19.7              \\ \hline
$1987$ & Bayes+MAR & 35.1 (34.0, 36.4) & 40.1 (38.3, 42.0) & 15.5 (14.8, 16.2) & 13.0 (12.0, 14.1) \\ 
$1987$ & Bayes+MNAR & 38.9 (36.2, 41.1) & 44.7 (41.8, 47.3) & 16.3 (15.2, 17.4) & 13.9 (12.6, 15.3) \\ 
$1987$ & Complete case & 34.8 (32.4, 37.2) & 39.8 (36.2, 43.4) & 15.4 (13.7, 17.2) & 12.9 (10.7, 15.3) \\ 
$1987$ & mice & 35.0 (32.5, 37.5) & 40.5 (36.9, 44.2) & 15.5 (13.8, 17.2) & 13.3 (10.9, 15.6) \\ 
$1987$ & True & 39.0              & 44.1              & 17.4              & 15.5              \\ \hline
$1992$ & Bayes+MAR & 35.0 (32.9, 37.1) & 37.7 (35.7, 39.6) & 12.7 (11.6, 13.9) & 15.0 (13.9, 16.2) \\ 
$1992$ & Bayes+MNAR & 32.6 (28.9, 37.2) & 36.0 (32.2, 40.1) & 15.6 (13.6, 17.8) & 16.7 (14.9, 18.9) \\ 
$1992$ & Complete case & 33.8 (30.3, 37.4) & 36.8 (33.2, 40.4) & 12.6 (10.3, 15.0) & 14.8 (12.4, 17.4) \\ 
$1992$ & mice & 35.2 (31.4, 39.0) & 38.0 (34.3, 41.8) & 12.7 (10.3, 15.0) & 15.2 (12.7, 17.8) \\ 
$1992$ & True & 38.0              & 39.8              & 16.0              & 17.4              \\ \hline
$1997$ & Bayes+MAR & 33.4 (31.7, 35.3) & 35.2 (33.1, 37.4) & 18.4 (17.2, 19.8) & 19.4 (18.1, 20.8) \\ 
$1997$ & Bayes+MNAR & 34.1 (29.7, 38.9) & 34.3 (29.8, 40.0) & 19.1 (16.7, 22.0) & 19.9 (17.8, 22.0) \\ 
$1997$ & Complete case & 32.6 (29.3, 36.0) & 34.0 (30.5, 37.7) & 17.9 (15.3, 20.6) & 18.9 (16.2, 21.7) \\ 
$1997$ & mice & 33.8 (30.3, 37.4) & 35.4 (31.5, 39.2) & 18.3 (15.9, 20.8) & 19.5 (16.8, 22.1) \\ 
$1997$ & True & 33.5              & 34.9              & 21.2              & 21.8              \\ \hline
$2002$ & Bayes+MAR & 35.2 (33.1, 37.2) & 44.2 (42.0, 46.3) & 22.2 (20.7, 23.7) & 17.9 (16.5, 19.3) \\ 
$2002$ & Bayes+MNAR & 36.2 (30.7, 42.9) & 44.2 (38.3, 50.2) & 22.8 (20.4, 25.5) & 17.2 (15.1, 19.6) \\ 
$2002$ & Complete case & 33.6 (30.1, 37.2) & 42.6 (38.8, 46.3) & 22.0 (19.1, 25.0) & 17.2 (14.6, 20.0) \\ 
$2002$ & mice & 35.2 (31.3, 39.1) & 44.2 (40.5, 48.0) & 22.1 (19.5, 24.8) & 18.1 (15.3, 20.8) \\ 
$2002$ & True & 35.3              & 44.3              & 25.0              & 19.1              \\ \hline
$2007$ & Bayes+MAR & 35.2 (32.7, 37.7) & 41.8 (39.2, 44.2) & 22.4 (20.6, 24.2) & 29.6 (27.7, 31.5) \\ 
$2007$ & Bayes+MNAR & 33.6 (27.7, 40.5) & 39.7 (33.0, 46.4) & 21.2 (18.2, 24.5) & 28.1 (24.7, 31.5) \\ 
$2007$ & Complete case & 33.3 (29.3, 37.5) & 39.9 (35.7, 44.2) & 21.7 (18.5, 25.1) & 28.9 (25.3, 32.7) \\ 
$2007$ & mice & 35.0 (30.6, 39.3) & 42.1 (37.8, 46.3) & 22.5 (19.3, 25.8) & 29.4 (26.1, 32.7) \\ 
$2007$ & True & 32.3              & 37.9              & 23.5              & 28.8              \\ 
\hline \\[-1.8ex]
	\end{tabular} }
\end{table}

\begin{table}[!htbp]
	\centering 
	\small
	\caption{Participant trends and model-based posterior trends with 95\% credible intervals for the FINRISK data.} 
	\label{table4} 
	\renewcommand{\arraystretch}{0.6}
	\resizebox{\textwidth}{!}{\begin{tabular}{@{\extracolsep{5pt}} clllll} 
		&  & \multicolumn{2}{c}{Northern Karelia} & \multicolumn{2}{c}{North Savonia} \\\hline
		year & method & men & women & men & women \\ 
		\hline \\[-1.7ex] 
		$1972$ & Bayes & 50.3 (49.0, 51.9) & 11.9 (11.0, 12.9) & 49.7 (48.6, 50.8) & 13.5 (12.7, 14.5) \\ 
		$1972$ & Complete case & 52.2 (50.3, 54.5) & 11.7 (11.0, 13.9) & 50.9 (49.4, 53.0) & 13.2 (12.0, 14.6) \\[2mm]
		$1977$ & Bayes & 48.5 (46.5, 51.0) & 13.2 (12.0, 14.5) & 46.4 (45.1, 47.9) & 14.4 (13.4, 15.4) \\ 
		$1977$ & Complete case & 43.1 (41.1, 45.4) & 8.9 (7.5, 10.4) & 43.1 (41.4, 45.0) & 11.0 (9.8, 12.3) \\ [2mm]
		$1982$ & Bayes & 43.9 (41.3, 46.5) & 18.9 (17.0, 21.1) & 46.5 (44.0, 49.0) & 18.8 (17.2, 20.6) \\ 
		$1982$ & Complete case & 36.1 (34.0, 39.1) & 14.2 (12.6, 15.9) & 42.6 (40.6, 45.6) & 15.9 (14.4, 18.2) \\ [2mm]
		$1987$ & Bayes & 39.0 (36.4, 42.5) & 17.8 (16.3, 19.6) & 43.6 (40.7, 46.7) & 17.2 (15.5, 19.2) \\ 
		$1987$ & Complete case & 34.4 (33.7, 38.9) & 15.7 (13.3, 16.6) & 39.8 (36.8, 43.6) & 15.3 (13.0, 17.7) \\ [2mm]
		$1992$ & Bayes & 35.5 (31.2, 39.8) & 17.6 (15.6, 20.0) & 38.3 (35.2, 41.7) & 20.1 (18.0, 22.1) \\ 
		$1992$ & Complete case & 31.2 (28.3, 36.1) & 16.0 (14.2, 19.2) & 35.3 (32.2, 39.7) & 18.0 (15.8, 20.7) \\ [2mm]
		$1997$ & Bayes & 33.5 (29.8, 37.5) & 20.2 (17.9, 22.9) & 32.5 (28.6, 36.8) & 19.5 (17.8, 21.5) \\ 
		$1997$ & Complete case & 31.1 (28.3, 35.4) & 16.5 (14.4, 19.6) & 30.7 (27.7, 35.5) & 17.0 (14.7, 19.6) \\ [2mm]
		$2002$ & Bayes & 34.5 (30.4, 39.2) & 24.8 (22.2, 27.6) & 36.5 (32.4, 41.1) & 21.3 (19.0, 24.1) \\ 
		$2002$ & Complete case & 32.5 (28.4, 35.8) & 22.4 (19.9, 25.4) & 34.3 (31.2, 38.6) & 19.2 (17.0, 22.1) \\ [2mm]
		$2007$ & Bayes & 30.4 (25.5, 35.5) & 18.6 (15.9, 21.7) & 29.3 (24.4, 35.5) & 20.4 (17.6, 23.5) \\ 
		$2007$ & Complete case & 29.2 (26.0, 34.0) & 17.3 (14.5, 20.5) & 28.3 (25.7, 34.1) & 19.5 (16.9, 23.0) \\
		\hline \\[-1.8ex] 
	\end{tabular}} 
\end{table}

\end{document}